# High-resolution and ultra-low power nonlinear image processing with passive high-quality factor metasurfaces

**Authors:** Bo Zhao[1], Lin Lin[1,2], Ameyaw Samuel[1], Mark Lawrence[1]*

**Affiliations:**

[1]Department of Electrical & Systems Engineering, Washington University in St. Louis; St. Louis, MO 63130, USA.

[2]Department of Chemistry, Washington University in St. Louis; St. Louis, MO 63130, USA.

*Corresponding author. Email: markl@wustl.edu

**Abstract:** Image processing is both one of the most exciting domains for applying artificial intelligence and the most computationally expensive. Nanostructured metasurfaces have opened the door to the ultimate energy saving by directly processing ambient image data via ultra-thin layers before detection. However, a key ingredient of universal computation - nonlinear thresholding functions - have yet to be demonstrated for low intensities without an external power source. Here, we present a passive, all-optical method for nonlinear image processing using Silicon nanoantenna arrays. We experimentally demonstrate an intensity thresholding filter capable of processing one-dimensional images with only Watt-level power. By leveraging the opto-thermal Kerr nonlinearity through high-Q guided mode resonance, we achieve an experimental threshold as low as 0.1 mW/μm² with a spatial resolution of 1.85 μm. Additional simulations indicate that the threshold can be further reduced while maintaining high spatial selectivity. Analog, pixel-wise, smoothed leaky ReLU activation filters promise to revolutionize image sensing.

## MAIN TEXT

### Introduction

The direct analogy between engineered optical structures and digital algorithms found ubiquitously throughout image processing software points to an opportunity for revolutionizing computational imaging and smart machine vision. Especially when aiming for sophisticated tasks such as superhuman self-driving cars and AI diagnostics, the possibility to circumvent the trade-off between precise performance and high energy consumption by replacing digital instructions with passive optical systems is exciting. All-optical filters have enabled almost any linear transformation on incoming images, including the Fourier transform [1–3], 1st and 2nd order differentiation, edge detection [4–6], various amplitude and phase convolutions, denoising, and object recognition [7–10]. Many useful operations have been realized for both coherent and incoherent illumination, across multiple wavelengths, and with polarization multiplexing [11–16], making them compatible with everyday light sources. However, linear transformations [1–3] represent just a tiny sliver of the space of universal computations. Hints that nonlinearity may hold the key to unlocking super charged all-optical image processing can be found in recent optical computing experiments [10,17–19]. Intensity dependent effects have produced unambiguous performance benefits over linear systems, but observing those effects either required huge light intensities, an electrical power source to provide extreme cooling or optical gain, and/or the sacrifice of spatial details [20–22].

The main obstacle standing in the way of passive low power optical nonlinearity is the lack of strong light-matter coupling in un-biased materials kept at room temperature. Weak material coefficients can be amplified within optically thick structures including long multimode fibers or vapor cells, but the function applied in these cases is highly non-local because mode mixing occurs faster than the nonlinear evolution [23–26]. Nonlinear amplification or absorption can be observed for very weak illumination after pumping amplifiers close to saturation, or cooling polaritonic microcavities or magneto-optically trapped Rb atoms below a few kelvins [27–29]. Similarly, diffractively coupled laser arrays operating close to the threshold can exhibit multimode nonlinear dynamics. Aside from requiring an external energy source, these schemes typically operate on a limited set of spatial modes. Circumventing the need for extreme material properties, various nonlinear encoding schemes have been proposed [30–32]. Unfortunately, devices based on this approach must program the input data into the control parameters instead of accepting raw images.

Nanostructured flat optical components known as metasurfaces have emerged as a way of locally enhancing optical nonlinearities by resonantly amplifying near fields [33–39]. They have been used to demonstrate many exciting nonlinear wave shaping phenomena [40–42]. However, metasurfaces are so thin that plasmonic and dielectric resonances fail to make up for the small third order susceptibilities of typical metals and semiconductors, constraining the Kerr-modulation efficiencies to well below that of macroscopic crystals. A recent article incorporates a phase change material into a nanostructured metasurface, displaying an all-optical limiting behavior at reduced illumination power [43]. While impressive, the suppression of high-power information is the opposite of the thresholding transformation we seek here and the use of $VO_2$ limits its optical transparency range. To further amplify the near field, several high-Q metasurfaces using high-order Mie resonance, quasi-bound states in the continuum [44–47] have been shown to interact with free-space light and enhance light-matter interactions to the level comparable to those achieved with ring resonators and photonic crystal defects. However, due to a trade-off between mode localization/angular dispersion and resonant enhancement, pixelated nonlinear image thresholding filters have not been realized yet.

Here, we experimentally demonstrate passive and all-optical nonlinear image processing of one-dimensional images containing Watt level powers. Specifically, our devices act as intensity high-pass filters, blocking incident light below some intensity threshold but becoming transparent above it. We achieve thresholds as low as 0.1 mW/μm² by resonantly enhancing the opto-thermal Kerr effect in silicon nanostructures. Allowing us to combine low threshold transparency with fine spatial selectivity corresponding to a resolution of 1.85 μm in the relevant dimension, we introduce metasurfaces supporting dipolar guided-mode resonances (DGMRs) with measured Q over 1000 and a non-dispersive angular range of 40°. We also show numerically that the same degree of locality can be maintained with significantly higher Q, and therefore a nonlinear response to much lower intensities should be possible. Nonlinearly processing high dimensional image data with a passive device offers an energy free route to dramatically expanding the representational capacity of all-optical neural networks [10,20,24,28,48–52].

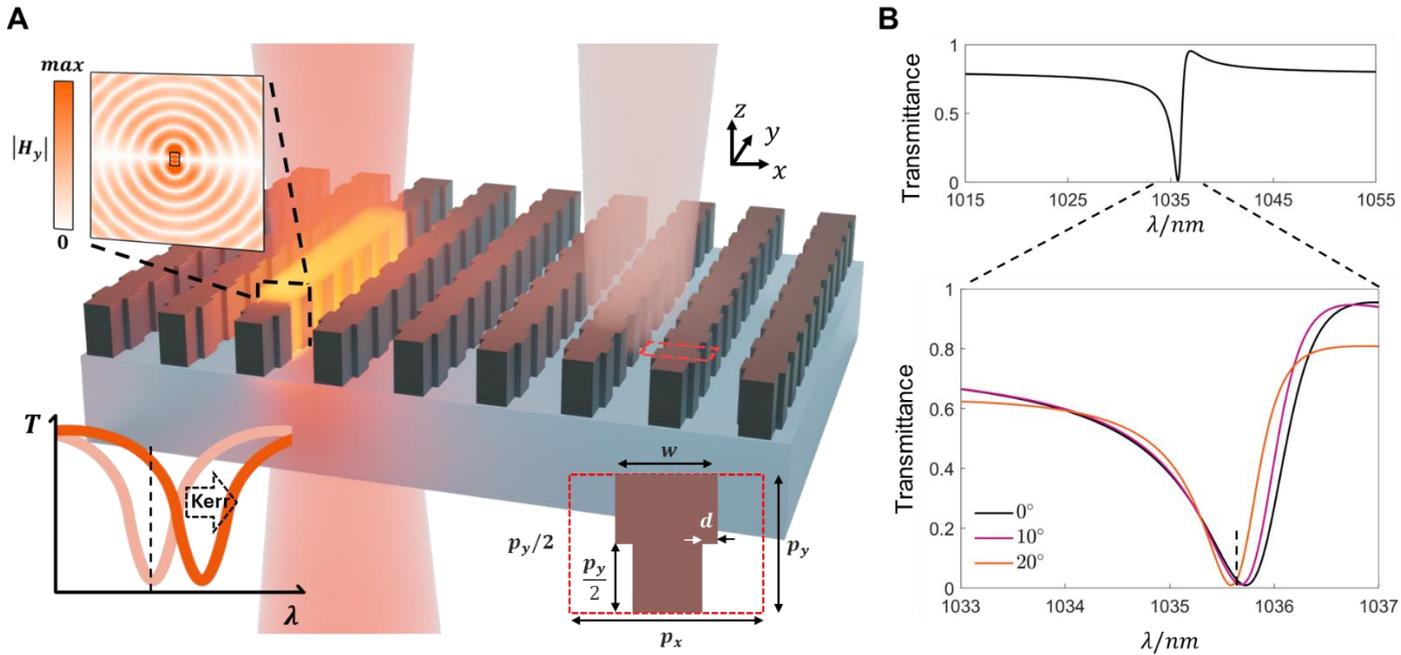

**Fig. 1. General operating principle of nonlinear meta-filter and simulated design of high-Q angularly non-dispersive DGMR metasurface.** (A), Schematic illustration of the designed metasurface, which transmits high-intensity spots while suppressing low-intensity spots. The intensity dependent transparency is produced by a redshift of the resonant transmittance dip mediated by the opto-thermal Kerr effect as shown in the bottom-left inset. Local transmission stems from each antenna supporting an independent DGMR that radiates isotropically, as seen in the top-left inset (simulated from a single antenna). Bottom-right inset: the top view of the unit cell design. (B), Top: Simulated transmission spectrum for the structure in panel (A) at normal incidence shows a high-Q, high contrast, Fano resonance. Bottom: Zoomed-in spectra of DGMR remain stable at different incident angles. The simulations is at the settings: $w_x = 360$ nm, $d = 120$ nm, $p_x = 505$ nm, $p_y = 355$ nm.

**Design of high resolution and low power nonlinear intensity thresholding metasurface**

A cavity supporting a spectrally isolated photonic resonance and exhibiting a corresponding transmission dip is a natural choice when building a transmissive intensity thresholding filter. Aligning the resonant wavelength to that of the illumination produces strong reflection and therefore efficient suppression of transmitted light. Increasing the refractive index of materials inside the cavity will cause the resonant wavelength to redshift, increasing the transmission until the film becomes almost fully transparent after a shift of roughly a linewidth. The key to connecting this transmission modulation to the amplitude of the incident wave is an effective intensity dependent refractive index, as shown schematically in the bottom-left inset of Fig. 1A. Not only is the opto-thermal Kerr effect one of the strongest Kerr-like nonlinearities it can also be enhanced by engineering straightforward linear light-matter interactions. The instantaneous nonlinear Kerr effect adds a term to the refractive index of a material equal to the product of a tiny constant and the local light intensity, so a Kerr-nonlinear shift of a resonator increases in proportion to its Q-factor thanks to a corresponding intensity amplification. Thermally mediated nonlinearity in a weakly absorbing cavity benefits from a very similar enhancement. When a material, in our case silicon, absorbs light, the material will heat up. Like most other semiconductors the refractive index of silicon increases linearly with temperature, so a resonant shift proportional to the incident optical power absorbed will be observed, describable in terms of an effective opto-thermal Kerr effect [53,54]. Given that silicon is reasonably sensitive to temperature, $\Delta n \sim 2e^{-4}$ $K^{-1}$, even a small amount of absorption will produce a change in index that is orders of magnitude larger than those of electronic origin, meaning the effective opto-thermal nonlinear coefficient can be much larger than $\chi^{(3)}$. In a cavity made from a material with a sufficiently small but finite absorption coefficient, the transmission spectrum mimics a lossless resonator while the fraction of absorbed power and therefore size of resonant shift will increase linearly with Q. At the same time, Q also defines the full width at half maximum (FWHM) of the transmittance dip. Shrinking the FWHM will therefore make the transmittance more sensitive to changes in resonant wavelength. Taken together, these two resonant enhancement mechanisms lead to an opto-thermal nonlinear threshold that scales with $Q^{-2}$. Counter intuitively, the efficiency of nonlinear transparency produced by a given resonator can be improved by boosting material absorption. One may worry that this would degrade the filtering performance, but in fact the transmission contrast between low and high power illumination remains insensitive to loss until the absorptance grows above about 40%. Therefore, the optimal Q occurs when the radiation loss is balanced with material loss. Different materials have different absorptivities, but the absorption rate of many materials also varies dramatically with wavelength. For example, the imaginary part of silicon's refractive index changes by more than three orders of magnitude between $\lambda$=0.5um and $\lambda$=1.1um. Working with crystalline silicon just beyond $\lambda$=1um, we find the optimum Q for realizing low-power and high contract nonlinear thresholding to be around a thousand.

To design a local metasurface that behaves as an array of pixelated intensity high-pass filters, we opt for meta-atoms supporting a new kind of photonic mode known as a dipolar guided mode resonance (DGMR) which confines light to two dimensionally sub-wavelength regions. An example of such a metasurface is illustrated in Fig.1A, consisting of 600-nm-tall silicon nano-antennas sitting on the sapphire substrate. In the absence of notches, each antenna would support a series of guided modes that remain perfectly bound because they possess larger momentum than free-space radiation. By breaking translational symmetry, introducing notches periodically allows light to couple in and out of the guided modes from free space. More specifically, for frequencies at which the gap period $p_y$ matches a guided mode wavelength, guided mode standing waves emerge which appear as scattering resonances. The free-space wavelength of a GMR can be controlled by adjusting $p_y$, while the Q and radiation loss are determined by the strength of translational symmetry breaking, which here is controlled by the notch depth $d$, achieving unlimited Q in theory and reaching up to several thousands in experiments. The question remains of how to obtain finest possible spatial discrimination. Instead of modeling the focusing of a Gaussian beam directly, we choose to exploit the duality between real and Fourier space. Specifically, we search for high angular independence and correspond it to high spatial resolution by applying the Rayleigh condition $R = 0.61 \lambda/\text{NA}$. The numerical aperture $\text{NA} = n \sin \theta$, where $\theta$ can be read from the maximum incident angle before the transmittance curve deviates half of FWHM. We opt to engineer a x-oriented electric DGMR. This DGMR generates an isotropic emission pattern as indicated by the left and top inset in Fig.1A. Under x-polarized illumination in Fig. 1B, the transmittance dip at $\lambda_1 = 1035.88$ nm shifts very little as the incident angle about y-axis increases up to $\theta = 20°$. This result therefore guarantees that our metasurface DGMR with $Q > 1000$ responds uniformly over a numerical aperture of 0.34 and spatial resolution of about 1.85 μm in the x-direction. The spatial resolution in y-direction is several tens of microns, relating to longitudinal localization of the GMR which can be estimated by comparing the guided mode group velocity with the resonant lifetime [55]. More importantly, the GMR can be further localized via photonic mirror at the end of antennas and promise NA larger than 0.1 in the y-direction [56]. Though the pixel is rectangular, such high resolution in nonlinear transmissive image processing is excellent compared to other power thresholding filters (see table 1) and still competitive with most existing linear counterparts [57,58]. Note that the angular independence still holds for even high Q design, and fig. S1 show an example with Q of 5136 at 1285.86nm. The residual angular dispersion with incident angle larger than 20° comes from the nearest neighbor coupling between nanoantennas. Boosting the Q enhances nonlinearity, but it also amplifies the impact of small overlapping evanescent fields between neighboring elements, producing nonlocal scattering. Increasing the width of antennas $w_x$ leads to better confinement of light and decreases the amplitude of evanescent fields, while increasing the spacing between them $s_x$ drops the overlapping of evanescent fields. We have minimized coupling by balancing $w_x$ and $s_x$, while maintaining the subwavelength limit by setting $p_x = w_x + s_x = 505 nm$. It deserves discussion that other modes with uniform radiation patterns, such as x- and y-oriented magnetic DGMR [59] supporting spherical radiation, can also be applied and extend the function to y-polarized incidence. But these are still sensitive to coupling. Other methods to eliminate the coupling, such as recent work from our group, building zero-coupling singularity [60] by destructively interfering longitudinal electric field with transverse electric field, are promising to solve this issue. Nevertheless, the spectra in Fig.1B have a slight asymmetry compared to ideal Lorentzian dips, which comes from the presence of a low frequency, low Q Mie resonance. It always requires some consideration of the background broadband Mie resonances, since their angular dispersion can also affect the curvature stability of DGMRs via interference.

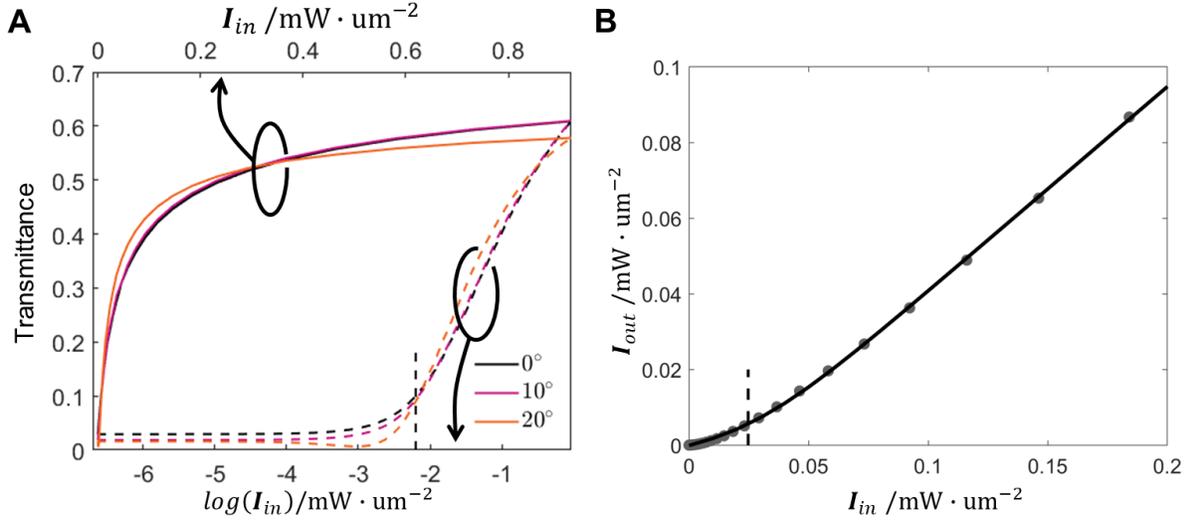

**Fig. 2. Simulations demonstrating optical thresholding from resonance enhanced optothermal Kerr nonlinearity.** (**A**), Simulated transmittance variation with input intensity (solid curves/ top x-axis) and log-scaled input intensity (dashed curves/ bottom x-axis) at different incident angles. Vertical dashed line: transition from linear to nonlinear response occurs at 0.0063 mW/μm². (**B**), Output versus input intensity response of the metasurface revealing a smoothed leaky ReLU function. Dashed line: fitted threshold of the smoothed leaky ReLU function at 0.0247 mW/μm² (see Methods for fitting details).

Unlike instantaneous mechanisms underpinning third order susceptibilities which depend on the electronic structure of atoms, opto-thermal nonlinearities can be engineered at the nano to microscale by controlling how heat dissipates from the nanostructures. To determine the nonlinear response of the metasurface, we use an analytical, 3-D conduction model to estimate the temperature increase and use the finite element method (see Methods for simulation details) to resolve the optical response. In the model, $\Delta T = Q_e/Sk$, where $Q_e = \int \frac{1}{2}\varepsilon_i |E|^2 \, dV$ is the total electromagnetic energy absorption and $\varepsilon_i$ is the imaginary part of silicon permittivity, $S = \frac{\pi a}{\ln(4a/b)}$ is the conduction shape factor and $k$ is the thermal conductivity of sapphire. The model corresponds to a heated rectangular plate on a semi-infinite medium, where the heated source is the illuminated region, and $a = 332\mu m$, $b = 130\mu m$, matching the laser spot size in the experiment. The refractive index increases linearly with respect to temperature as $n_r = n_{r0} + \alpha \Delta T$, where $n_{r0}$ is the refractive index at room temperature and $\alpha = 2.1892 \times 10^{-4} \text{ K}^{-1}$, following Jellison's model [53]. Simultaneously solving for the thermal and optical steady states, Fig. 2A demonstrates that illuminated by x-polarized light at $\lambda = 1035.6nm$, the transmittance increases with respect to irradiance, shown in both linear (solid) and log (dashed) scales, similarly at all the incident angles. Most excitingly, nonlinear behavior is seen to set in at the incredibly low intensity of 0.0063 mW/μm², which is distinct from low Q thermo-optic approaches which are dominated by absorption loss [61,62]. Note that the transmittance contrast can be pushed towards 100% and the rate at which the transition from nominally opaque to transparent occurs can be improved by designing the spectral line shape to have the opposite Fano factor, similar to Fig. 1B. From the input/output curve in Fig. 2B, the metasurface filters the light as a smooth approximation to the leaky rectifier (smoothed leaky ReLU) which is frequently used in digital neural networks but has yet to be applied in passive optical computing. The gradient of the fitted smoothed leaky ReLU is a Sigmoid function, smoothly changing from the lower boundary of 0.07 to the upper boundary 0.54 in the fitting range of 0 to $0.2 mW/\mu m^2$. The threshold of the smoothed leaky ReLU function sets as the middle point of this changes and is as low as $0.0247 mW/\mu m^2$ (see method for details). The finite element simulation results in fig. S2 and fig. S3 illustrate how heat originating from a central nanoantenna spreads through the rest of the metasurface. Due to the strong thermal isolation provided by the air, the rise in temperature of the nearest neighbor antennas is 59% cooler than the central antenna under continuous wave illumination and much larger with a pulsed laser. This guarantees that the thermal dissipation length is smaller than the optical spatial resolution. The thermal isolation of the nanoantennas can be further improved via over etching trenches in the substrate (see details in fig. S3), and the threshold can also be further reduced by designing a higher Q optical response.

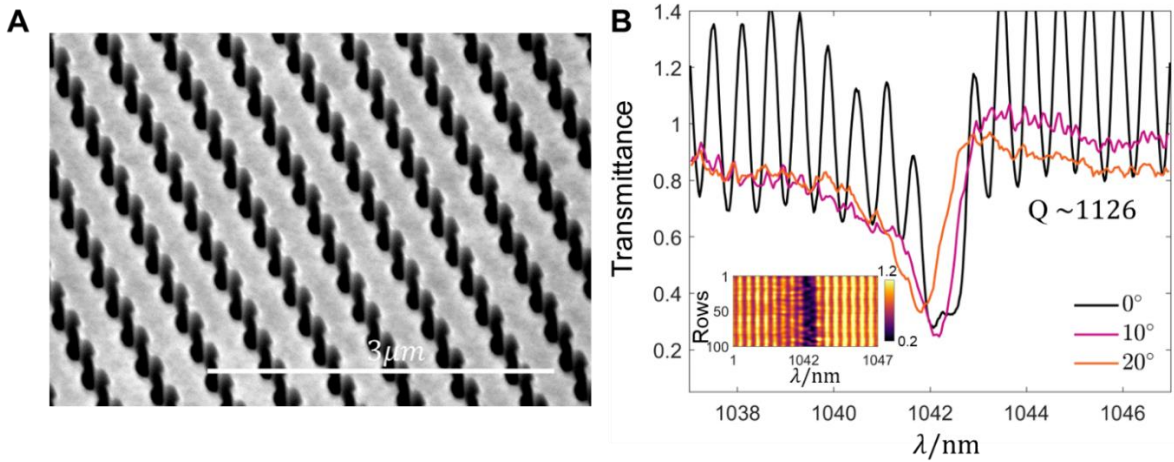

**Fig. 3. Experimental demonstration of high-Q angular non-dispersive metasurface. (A),** SEM image of the fabricated metasurface based on the design in Fig. 2. **(B),** Measured transmittance spectra at different incident angles. The transmittance is normalized by light passing through the non-patterned area and averaged over about 100 antennas. The fitted Q is 1126±261, averaged from these rows (see fig. S4 for fitting details).

**Measurement of angular independence and smoothed leaky ReLU nonlinearity of the metasurface**

Based on the design in Fig. 2, we have fabricated various metasurfaces, an SEM image of an example of which is shown in Fig. 3A. After polarizing a supercontinuum laser to match the orientation of the DGMR, we measured the transmission spectra of each metasurface with an imaging spectrometer. The samples are also mounted on a rotation stage, allowing the incident angle to be varied. The measured transmittance curves in Fig. 3B reveal a resonance dip close to a wavelength of $1042nm$ with a Q of 1126. Fig. 3B also shows that this sample displays minimal incident angle dispersion, with the resonance shift staying within a FWHM up to 20º, which fits well with the simulation. This confirms that the structure is responding with a high degree of spatial locality. We note that the fringes uniformly spread across all measured spectra come not from noise but instead from Fabry-Perot interference in the thick substrate. Evidence for this is seen in our ability to accurately represent both the resonant dip and background oscillations when curve fitting in fig. S4D. The spectral image for normal incidence given in the inset of Fig. 3B, as well as those for the other angles in fig. S4 of the supporting information, also demonstrates that the sample is highly uniform.

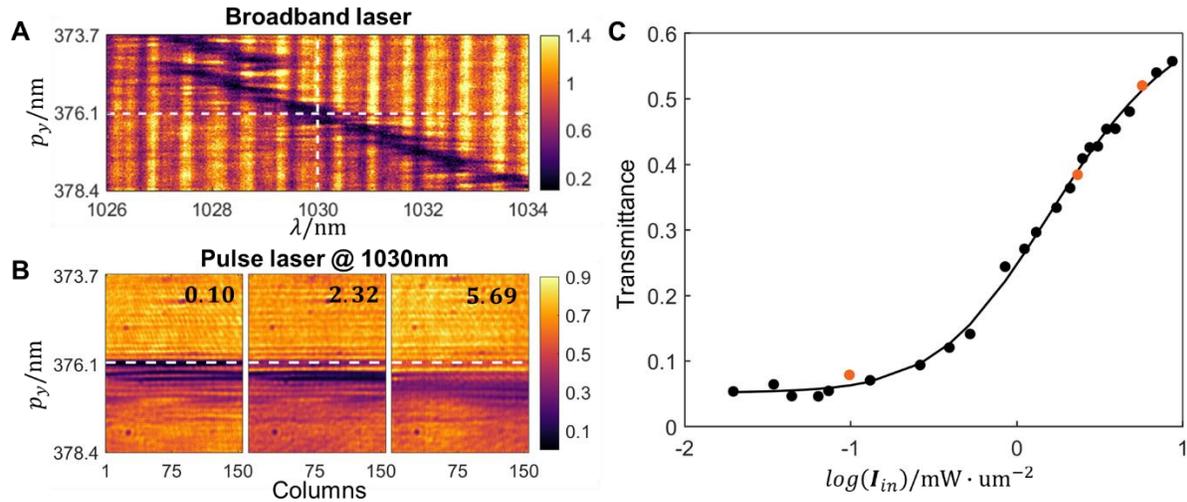

**Fig. 4. Observation of nonlinear transparency with ultra-low intensity.** (**A**), Measured spectral image of a metasurface consisting of period-swept nanoantennas, revealing a corresponding variation in resonant wavelength. The dashed white lines mark the period at which the resonant dip perfectly matches the wavelength of the pulse laser. The $y$-axis corresponds to the vertical pixel number on the camera (see fig. S6 for details), the label for which has been converted to nanoantenna period. (**B**) and (**C**) reveal the nonlinear response. (**B**) shows the spatially dependent transmission through the period-swept metasurface of the λ=1030nm pulsed laser after attenuating to three different powers. The input intensity $I_{in}/mW \cdot um^{-2}$ is labeled in the corner. (**C**), Transmittance as a function of input intensity for the metasurface region with resonance matching the laser wavelength (dashed line in (**B**)), where the orange dots represent the data from (**B**).

To measure the nonlinear properties of our metasurfaces, we switch to a $0.5ns$ pulsed laser with a fixed wavelength of 1030 nm. To align the resonant wavelength with that of the laser, we fabricate an inhomogeneous metasurface consisting of 801 nanoantennas with notch period $p_y$ swept continuously from 372 nm to 392 nm. The corresponding transmittance spectral map of the metasurface under broadband illumination is shown in Fig. 4A, where the resonance redshifts continuously with increasing $p_y$. When shining the pulsed laser on this sample after heavy attenuation, it can be seen in Fig. 4B that transmission is fully blocked only near the region of the metasurface supporting a resonance matching up with the laser wavelength, highlighted by a dashed white line, which produces a spatially localized dip. Although unsmooth defects from fabrications have produced some other dim resonant regions in Fig. 4B, they are far enough from the interest region to not influence its thermal dissipation. Next, we increase the average illumination intensity $I_{in}$, defined as the measured peak power divided by the focused spot size, and record the transmittance of the resonant region around the dashed line in Fig. 4B. As shown in Fig. 4C, the transmittance remains below 0.1 when $I_{in}$ is down in the µW/µm² regime. As $I_{in}$ increases, the transmittance begins to grow before saturating when the redshift approaches the initial linewidth. We emphasize here the threshold intensity separating linear and nonlinear behavior is only 0.1 mW/µm², which is 2 to 3 orders of magnitude lower than any previous point-wise thresholding structure [21]. In table 1 our result is also compared to other power thresholding and limiting devices operating on free space light waves. It is worth mentioning that the repetition rate of the pulsed laser we use is 10 kHz, but the metasurface cools down after heating by 500 ps long pulse in about 32 ns (see details in fig. S2). We therefore expect the power required to induce a nonlinear response can be reduced by two orders of magnitude, corresponding to sub-Watt total powers, using a continuous wave laser.

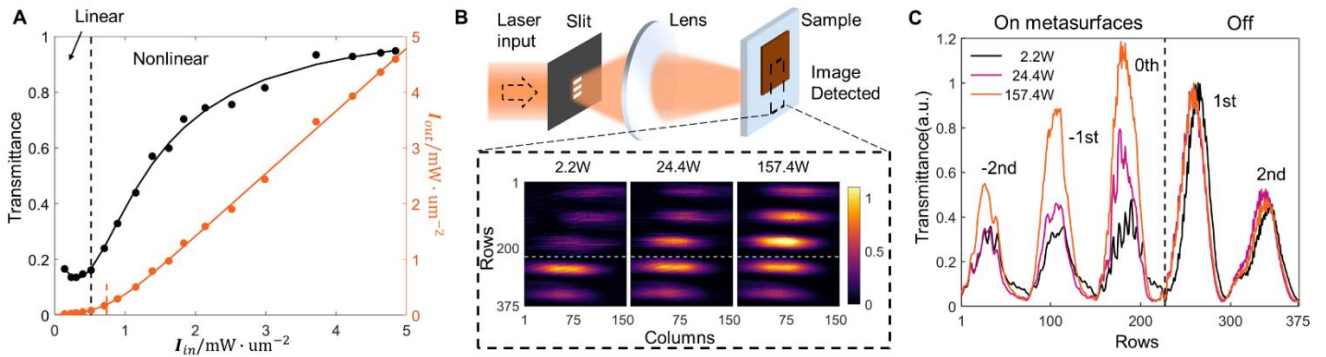

**Fig. 5 Experimental demonstration of nonlinear image filtering.** (**A**), Measured (black dots) and fitted (black curve, see fitting details in Methods) nonlinear response of the uniform metasurface. The transition intensity from linear to nonlinear response is indicated by dashed black line at 0.52 mW/µm². Recorded output-versus-input intensity response (orange dots) and fitted smoothed leaky ReLU function (orange curve), which yields a threshold of 0.738 mW/µm² as the dashed orange line. (**B**), Top: experimental setup, including a test object consisting of slits engraved into a metal foil. The diffraction pattern from this object is projected onto the metasurface sample. Bottom: filtered images at different total incident powers. The images are normalized by the 1st order diffraction peak and the total power is recorded by a thermal detector before the sample. (**C**), Transmitted power distribution along rows, with the power averaged across all the columns.

### Filtering a diffraction pattern

We then fabricate a series of uniform samples, each with a slightly different period in the y direction, before selecting the resonance with $p_y = 376 nm$ that matches the laser most closely.

The nominal nonlinear filter function for the chosen metasurface is shown in Fig. 5A, which is collected from the average over nearly 100 antennas. The power dependence approximates a smoothed leaky ReLU function with a threshold of 0.738 mW/μm². Since the resonant wavelength is slightly blue-shifted relative to the laser, the transmittance is seen to drop slightly before rising with increasing input power. This causes the threshold intensity to be a little bit higher than Fig. 4 at around 0.52 mW/μm². Based on the large numerical aperture, we are confident to input an image. To generate a 1D test image, we have cut three slits from an aluminum foil screen. Placing the screen in front of the laser, we project the Fourier image of the transmitted diffraction pattern to the sample via a cylindrical lens. Three slit diffraction contains several spots of varying amplitudes (only 5 main peaks from -2nd to 2nd orders are shown here) as shown in the top of Fig. 5B. To calibrate the nonlinear response, we positioned the metasurface so as to filter only the 0th, -1st and -2nd order peaks, while the 1st and 2nd order peaks bypass the metasurface and act as a reference. In Figs. 5B-C, the transmitted patterns for three different total powers, measured by focusing the diffraction after the lens to a power meter, are compared by normalizing the 1st order peak in each pattern.

Averaging the transmitted power along the nanoantenna direction, Fig. 5C plots the metasurface transformed distributions of the 1D diffraction profile. When the total illumination power is kept below 2.2 W, all of the transmitted peaks passing through the metasurface are suppressed but not totally dark since the transmittance minimum is low but not zero. Crucially, as the illumination becomes brighter, not only does the filtered pattern get brighter, but the shape of the pattern evolves. Since the 0th order peak has the largest relative intensity, it is the first to experience a nonlinear response, becoming amplified with respect to the -1st/-2nd peaks as shown in Fig. 5B with a total power of just 24.4 W. Further increasing the total power to 157.4 W, the intensities of all of the peaks are seen to become large enough to saturate the metasurface, causing the transmitted distribution to revert back to the unfiltered diffraction pattern. Considering that half of the visible peaks are outside of the metasurface and the total area of the diffraction pattern is much larger than that shown in Fig. 5B, after focusing the pattern to better match the meta-filter we expect to observe nonlinear filtering with <10 W. Also, we note that the transmitted pattern is not perfect due to the nonuniformity from fabrication as shown in fig. S5. The resonance at the region of the -2nd peak is slightly blue shifted and so transparency sets in at a higher intensity, causing the change in the amplitude of the -2nd peak when increasing the total power from 2.2 W to 24 W to be less noticeable than the other peaks. However, these results are clearly consistent with a spatially local nonlinearity in which the transmission through some part of the metasurface depends only on the intensity hitting that part of the metasurface. To provide further evidence of locality, we analyze the shape of the 0th order peak by fitting this peak with a Gaussian function modified with the thresholding filter in Fig. 5A. The result in fig. S5D shows that the width of the peak gradually decreases before the nonlinearity is saturated. This is consistent with the central region having higher intensity than the sides and therefore triggering a larger change in transmittance as the total power increases, giving a clear signature of a local nonlinear response. These results convince us that we have realized a high-resolution super-low intensity threshold image filter.

| | Function | Materials @ Working wavelength | Optical Mode | Illumination Power (kW/cm^2) Pulse Laser [Duration time (ps)/Repetition Rate (kHz)] | Illumination Power (kW/cm^2) Continuous Wavelength laser |
|---|---|---|---|---|---|
| This work | Power Thresholding | Silicon-Sapphire @1030nm | GMR | 10 [500/10] | 0.2* |
| Cotrufo, M., et al. | Power Thresholding | Silicon-Silica @1550nm | Q-BICs | $3.5 \times 10^4$ [2/80] | |
| Guo, J., et al | Power Thresholding | Gold-Graphene-CaF2 @10.6um | LSPs | | $2.5 \times 10^3$* |
| Tripathi, A., et al | Power Limiting | PMMA-Silicon-VO2-Silica @1470nm | Mie Resonance | 22.3 [100/1] | 0.2 |
| Howes, A., et al | Power Limiting | Si-VO2 @1240nm | Huygens' Resonance at ENZ point | $5 \times 10^3$ [100]* | |
| Nishida, K., et al | Power Limiting | Si SiO2 @561nm | Mie Resonance** | | 60 |

Table 1. Comparison between this work and other image power thresholding/limiting filters

The cells are blank if the references have not recorded corresponding data. *Estimated results from simulation. Our simulation is shown in fig. S2. **This work functions as scatter limiting with low contrast.

**Discussion**

In conclusion, we have introduced and experimentally demonstrated a fully passive scheme for applying a spatially selective intensity threshold, approximating a smoothed leaky ReLU function, to images containing Watt-level powers. To access the extreme nonlinearity required to turn opaque nanoscale films transparent with illumination intensity as low as 0.1 mW/μm², we harnessed high-Q dipolar guided mode resonances to both boost absorption-based photo-thermal refraction and generate strong transmission sensitivity to index changes. Crucially, the resonant nanoantennas are engineered to be cross-sectionally local, producing an array of independent micrometer scale thresholding pixels. Having numerically confirmed that a metasurface with a 5× higher Q-factor can exhibit the same degree of locality as our experiment, we expect simple fabrication improvements can lead to a significant drop in threshold. Moreover, there is a lot of room to further engineer the thermal dissipation, potentially increasing the rate of index change with incident power by more than an order of magnitude. Our findings therefore point to the possibility of directly processing ambient images produced with cheap laser diodes and containing just a few milliwatts of power, representing a significant advance over existing all-optical image processing techniques. Responding on nano to micro-second time scales, the optothermal nonlinearity is not well suited to optical processing of digitally programmed data, but it is easily compatible with refresh rates of existing image sensors. In the current project, we opted to work with silicon at a wavelength close to its band edge, providing just enough absorption to efficiently generate heat, but not so much as to spoil a resonance with Q-factor ~1000. This nonlinear image filtering scheme can be readily adapted to other wavelengths by choosing/combining materials to balance the absorption loss and transmittance contrast. The design principle we have presented holds great versatility for integration with digital cameras or diffractive optical devices, offering high-resolution nonlinear image processing without consuming additional energy. We believe this development will bring exciting new capabilities to current low-cost image sensors.

**Materials and Methods**

Simulation

We use commercial finite element method software COMSOL6.1 to simulate the model. We simulate a unit cell with periodic boundaries in the x/y directions and input/output light by two ports in the z direction. In Fig. 1, the silicon stands on a sapphire substrate. The dispersive permittivity of silicon is set using Green's model [63] and $\varepsilon_{r\_sub} = 1.75^2$ is used for the sapphire substrate. When simulating the Opto-thermal Kerr effect, we add in the silicon permittivity an extra linear term with respect to temperature based on Jellison's model [53].

Fabrication

The metasurfaces were fabricated using standard lithographic procedures. The Silicon-on-Sapphire wafers (University wafers) were cleaned via sonication in acetone and isopropyl alcohol respectively. Following the dehydration nitrogen blowing, the wafers were further cleaned by Oxygen Etcher (Asher) and rinsed with isopropyl alcohol to restore the hydrophilic base. Positive-tone resist ZEP was spin-coated to the sample and baked for 3min at 180°C. The patterns were written using electron-beam lithography (Elionix ELS-S50EX) and developed in ZED-N50. We then used reactive ion etching (etched by $SF_6$ via Oxford Plasma Lab 100) to transfer the pattern to the silicon layer. The residual resist was stripped by PG remover solution heated to 80°C.

Optical Characterization and Fitting Details

The sample is mounted on a rotational stage in a home-built microscope setup (see details in fig. S6), allowing the linearly polarized light source to alternate between broadband laser (NKT supercontinuum) and $1030nm$ pulse laser (Thorlabs QSL103A). To observe the normalized transmittance, we record the transmitted light both on and off the metasurface at each incident angle. During the nonlinear measurement, the pulse laser is attenuated using a combination of polarizers and neutral density (ND) filters, with the input power intensity recorded by a thermal detector (Thorlabs S405C).

The resonant spectral features were analyzed by fitting the transmittance spectra to the function:

$$T_1(f) = \left|\frac{1}{1 + F \sin^2 n_{sub} k h_{sub}}\right| \left|a_r + a_i i + \frac{b}{f - f_0 + \gamma i}\right|^2$$

Here, the second multiplicative term represents the superposition between a constant complex background, $a_r + a_i i$, and a Lorentzian resonance with resonant frequency $f_0$ and full width at half-maximum $2\gamma$. The Q-factor of the resonance is then taken to be $Q = f_0/2\gamma$. The first term accounts for the Fabry-Perot resonance through the substrate, where $F$ represents the reflectivity of the air-substrate interfaces, $k$ is the wavevector of the incident wave, $n_{sub}$ is the substrate's refractive index and $h_{sub}$ is the thickness. For simulated data, this term is simplified to 1.

In nonlinear response fitting (Fig. 5A), the Fabry-Perot resonance term is also neglected due to the measurement being performed at a single wavelength. The resonant wavelength is assumed to be linearly proportional to the input intensity, and the transmittance data is fitted by:

$$T_2(I) = \left| a_r + a_i i + \frac{b}{\alpha I - \alpha I_0 + \gamma i} \right|^2$$

To analyze the nonlinear Output-versus-Input data, we fit the function:

$$I_{out} = \beta_y \cdot f(\beta_x(I_{in} - I_{th}))$$
$$f(x) = \alpha x + (1 - \alpha) \log(1 + e^x)$$

Here, $I_{in}$ maps $I_{out}$ via a scaled version of the smoothed leaky ReLU function $f(x)$, where the $\beta_x, \beta_y$ are scaling coefficients of x- and y- coordinates, and $I_{th}$ is the threshold intensity. The gradient of $f(x)$ is a modified Sigmoid function, smoothly changing from α to 1. Correspondingly, the fitted transmittance (the gradient of $I_{out}$ versus $I_{in}$) changes from $\beta_x\beta_y\alpha$ to $\beta_x\beta_y$. And the threshold informs the middle point of this change.

In fig. S5D, the shape of the 0th peak at 2.2W is fitted as a nonlinear thresholded Gaussian function:

$$I_{out}(x) = T_2(I_{in}) \cdot I_{in}$$
$$I_{in} = A e^{-\frac{(x-x_0)^2}{2\sigma^2}}$$

The parameters for $T_2(I)$ are consistent with those in Fig. 5A, and the fitted $\sigma = 14$ matches the peak width. To illustrate the local effect of nonlinear thresholding, only the amplitude $A$, corresponding to the recorded power, is adjusted and the normalized filtered peak is plotted. The result shows that at higher power, the peak becomes narrower, consistent with predictions from the formula.

## Acknowledgments

**Funding:** National Science Foundation (NSF) grant no. CCF-2416375 and Optica Foundation.

**Author contributions:** M.L. conceived the idea and supervised the project. B.Z. performed the numerical simulations and analysis. B.Z., L.L. fabricated the samples. B.Z., S.A. conducted optical measurements. B.Z., M.L. wrote the manuscript. All authors discussed the results and commented on the manuscript.

**Competing interests:** Authors declare that they have no competing interests.

**Data and materials availability:** All data needed to evaluate the conclusions in the paper are present in the main text and supplementary materials. The data and code that support the plots within this paper and other findings of this study are available from the corresponding author upon reasonable request.

**Correspondence and requests for materials** should be addressed to Mark Lawrence.

**Supplementary Materials**

Figs. S1 to S6

# Supplementary Materials for

**High-resolution and ultra-low power nonlinear image processing with passive high-quality factor metasurfaces**

Bo Zhao *et al.*

*Corresponding author. Email: markl@wustl.edu

**This PDF file includes:**

Figs. S1 to S6

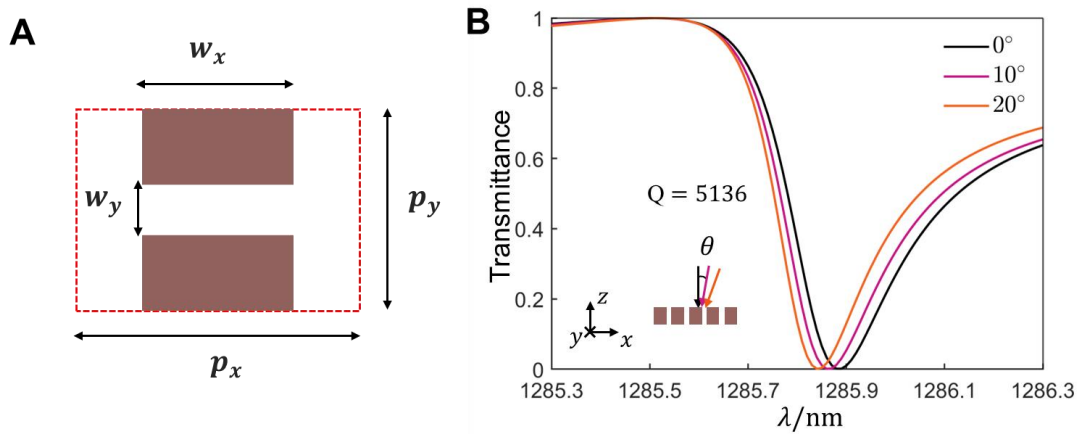

**Fig. S1. An example of angular independent metasurface with Q over 5000**
(**A**), The top view of a unit cell. The notches of antennas are cut down to maintain high transmission background at larger working wavelength. (**B**), Simulated transmittance spectra at different incident angles. The metasurfaces is simply suspended in the air. The silicon is lossless and non-dispersive, set as $\varepsilon_r = 3.7^2$. Such a choice is achievable by using polycrystalline and helps reduce near field coupling.

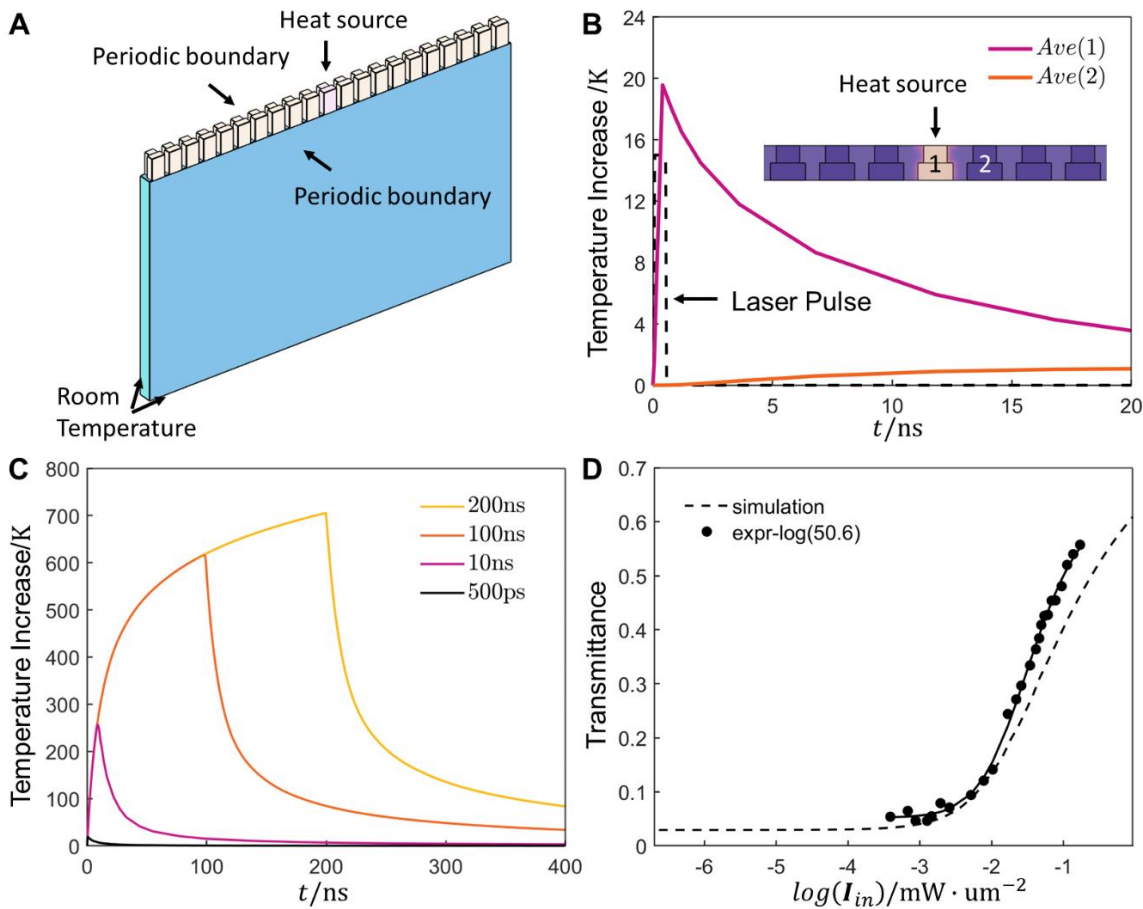

**Fig. S2. Thermal dissipation of the metaantenna under pulse laser**

(**A**), Simulation setup showing a single antenna dissipating heat to its surroundings. The central unit is modeled as a heat source with an energy density represented by a rectangular pulse with a 500 ps duration. The peak energy density is consistent with the energy absorption density calculated in optical simulation. (**B**), Simulation results show a temperature increase of 18K. The inset displays a top view of temperature distribution, highlighting the central antenna (heat source, labeled as antenna 1) and a neighboring antenna (antenna 2). The average temperature of antenna 1 increase up to 18K but that of antenna 2 maintain less than 2K during pulse excitation, showing that the thermal dissipation is well localized in a single unit, which is smaller than optical resolution. (**C**), The stationary temperature of the central unit increases and saturates at 987 K as the pulse duration increases, yielding a ratio of 50.6 between the simulation and experiment data. (**D**), Experimental data fits well with the simulation after applying a calibration of -log(50.6) to the x-axis.

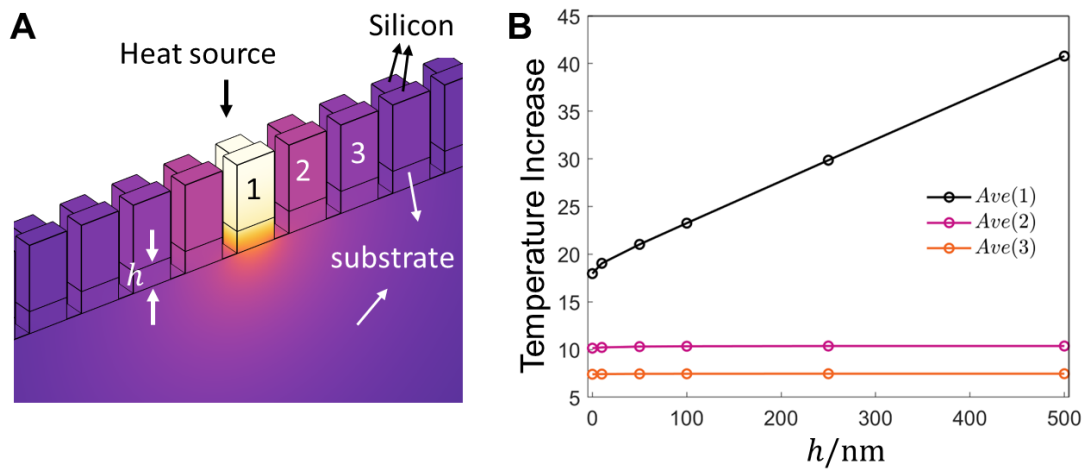

**Fig. S3. Effect of trenches on isolating heat dissipation**

(**A**), Schematic of heat dissipation in a structure with trenches. The trenches are etched into the substrate, following the shape of silicon antennas after over-etching. The central antenna (heat source) and a neighboring antenna are labeled as 1 and 2. (**B**), The averaged temperature increase in antenna 1 and 2 from (**A**) rises as the trench depth increases.

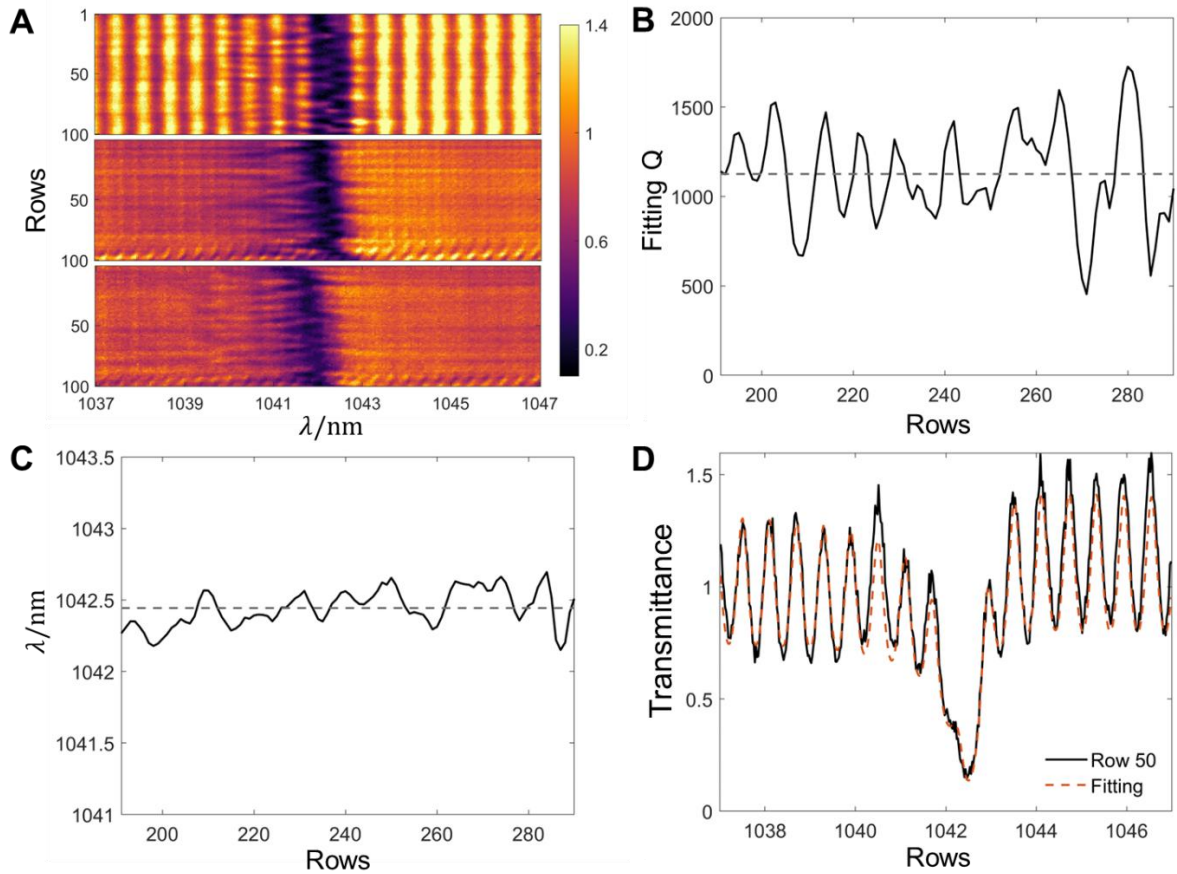

**Fig. S4. Measurement of angular independence and Q-fitting details**

(**A**), 2D spectra of the metasurface, with the y-axis representing the number of pixel rows from the camera, oriented along vertical-to-antenna direction. The inset in Fig. 3B corresponds to the top figure. (**B**), The fitting Q over 100 rows. The averaged Q is 1126 with the root-mean-square deviation of 261. (**C**), The fitting resonant wavelength versus different row numbers, with the average of 1042.4nm and the root-mean-square deviation of 0.13nm. (**D**), A fitting example of (**B**) and (**C**), which is recorded at the central row. The overlap between fitting curve and raw data indicates our fitting method successfully presents the Fano shape high Q GMR and Fabry-Perot resonance from the substrate.

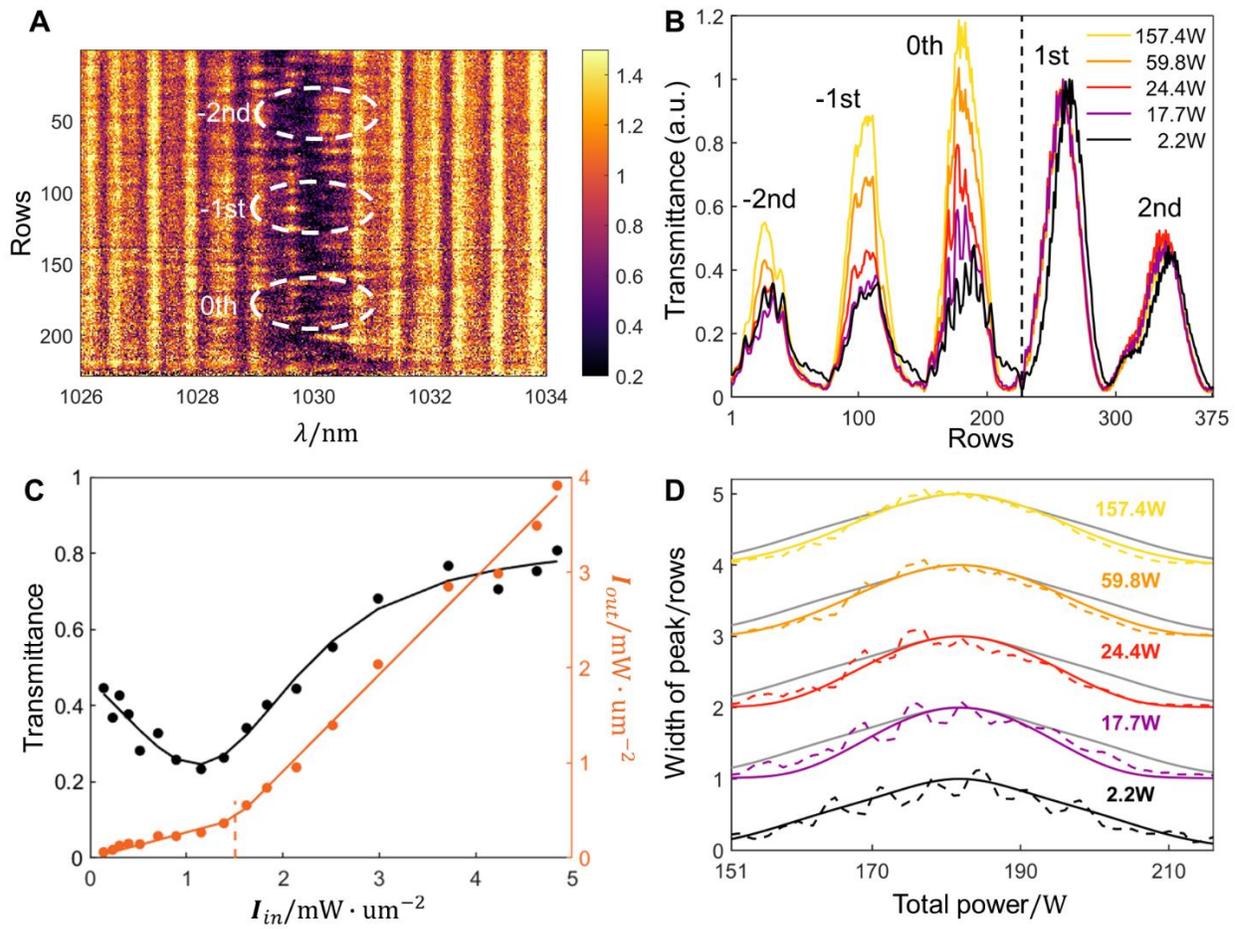

**Fig. S5. Details of image filtering**

(**A**), 2D transmission spectrum of the sample used for image filtering in Fig. 5, with the diffraction peaks highlighted by the dashed circles regions. (**B**), Transmitted power distribution at different powers. The relative transmitted power of the -2nd peak decreases as the total power increases to 17.7 mW. (**C**), Measured data, averaged over the -2nd peak region, shows a delayed nonlinear response. As the input power increases, the transmittance drops before reaching a dip, and the leaky ReLU threshold is higher than that of the -1st peak region. (**D**), The normalized shape of the 0th peak (dashed curve) and nonlinear filtered Gaussian prediction (solid curves) using the fitting method mentioned earlier. Data for different powers are vertically shifted for better comparison. The raw data at 2.2 W is shifted by -4 rows to correct for experimental misalignment. Gray curves represent duplications of the fitted curve at 2.2 W, emphasizing the shape narrowing effect due to nonlinear filtering.

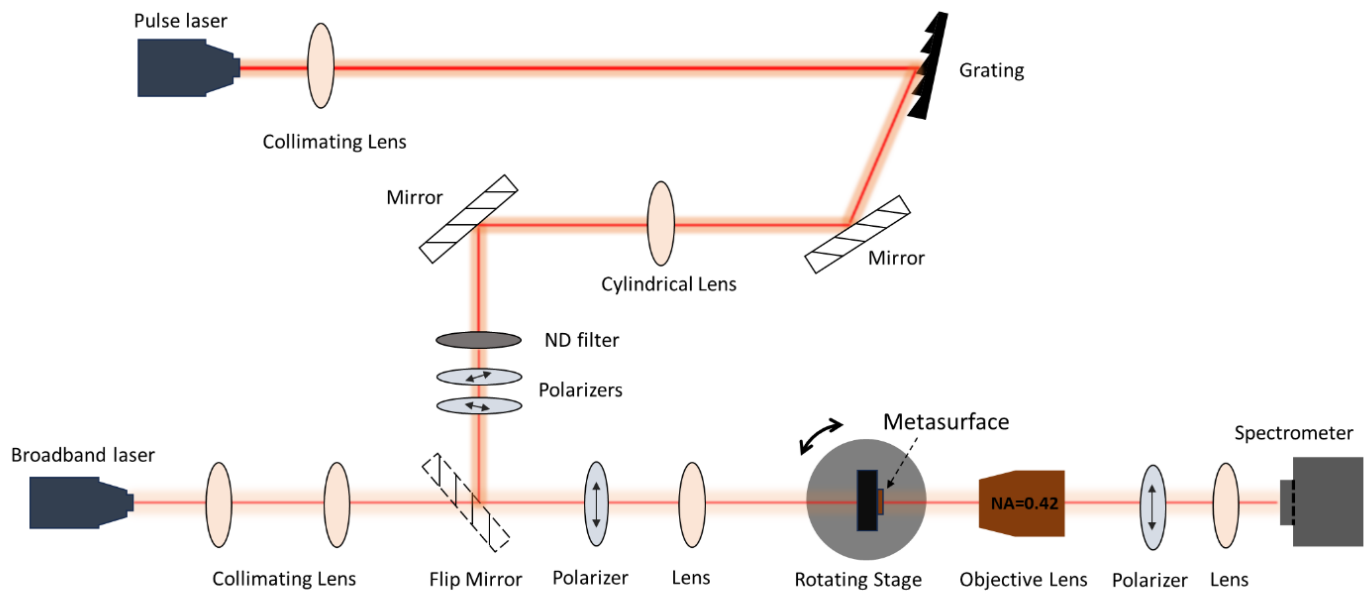

**Fig. S6. Schematic of a home-built microscope setup for optical measurement**